\documentclass{pasj00}
\draft
\begin{document}
\SetRunningHead{Yahata et al.}{FIR emission from SDSS galaxies}
\Received{2006/00/00}%{yyyy/mm/dd}
\Accepted{2006/00/00}%{yyyy/mm/dd}

\title{The effect of FIR emission from SDSS galaxies \\
on the SFD Galactic extinction map}

%%% begin:list of authors
%%% Please use the following style in case that sorting by 
%%% affilation is impossible. 

\author{%
   Kazuhiro \textsc{Yahata}\altaffilmark{1},
   Atsunori \textsc{Yonehara}\altaffilmark{1},
   Yasushi  \textsc{Suto}\altaffilmark{1}, \\
   Edwin L. \textsc{Turner}\altaffilmark{2},
   Tom      \textsc{Broadhurst}\altaffilmark{3}, 
    and
   Douglas P. \textsc{Finkbeiner}\altaffilmark{2}}
 \altaffiltext{1}{Department of Physics, School of Science,
The University of Tokyo, Tokyo 113-0033}
 \altaffiltext{2}
 {Princeton University Observatory, Peyton Hall, Princeton, NJ 08544,  
USA}
 \altaffiltext{3}{Department of Astronomy and Astrophysics, Tel-Aviv University, Israel}
 \email{yahata@utap.phys.s.u-tokyo.ac.jp}

%% `\KeyWords{}' always has to be placed before `\maketitle'.
 %Do NOT move this preamble from here!
\maketitle
%%%%%%%%%%%%%%%%%%%%%%%%%%%%%%%%%%%%%%%%%%%%%%%%%%%%%%%%
\begin{abstract}
We compare the most successful and widely used map of Galactic dust
extinction, provided by Schlegel, Finkbeiner \& Davis (1998; hereafter
SFD), to the galaxy number counts in the Sloan Digital Sky Survey (SDSS)
photometric/spectroscopic DR4 sample.  We divide the SDSS survey area
into 69 disjoint subregions according to the dust extinction provided by
SFD and compare the surface number density of galaxies in each
subregion.  As expected, the galaxy surface number density decreases
with increasing extinction but only for SFD extinction values above
about 0.1 to 0.2 magnitudes (depending on the band).  At lower values of
the SFD extinction, we find that {\it the sky surface density of
galaxies increases with increasing extinction, precisely the opposite of
the effect expected from Galactic dust.}  We also find that the average
color of the SDSS photometric galaxy sample is bluer at higher SFD
extinctions in this regime, again the opposite of the effect expected
from Galactic dust.  Even though these anomalies occur only for
sight-lines with low SFD extinction values, they affect over 70\% of the
high Galactic latitude sky in which galaxies and their clustering
properties are normally studied.  Although it would be possible to
explain these effects with a mysterious component of Galactic dust which
is anti-correlated with the 100$\mu m$ flux on which the SFD extinction
map is based, this model is not physically plausible.  Moreover, we find
that the surface number density of SDSS photometric quasars does not
show any similar effect, as would be expected if the explanation were an
unknown Galactic dust component.  Considering these results, we suggest
that the far infrared (FIR) brightness of the sky in regions of true low
dust extinction is significantly ``contaminated'' by the FIR emission
from background galaxies.  We show that such an explanation is both
qualitatively and quantitatively consistent with the available data.
Based on this interpretation we conclude that systematic errors in the
SFD extinction map due to extragalactic FIR emission are quite small, of
order hundredths of a magnitude, but nevertheless statistically
detectable.  Unfortunately, however, these errors are also entangled in
a complex way with a signal of great interest to many ``precision
cosmology'' applications, namely the large scale clustering of galaxies.
\end{abstract}
%%%%%%%%%%%%%%%%%%%%%%%%%%%%%%%%%%%%%%%%%%%%%%%%%%%%%%%%%%%%%%%%%%%
\section{Introduction}

Understanding the properties of dust is a fundamental goal in astronomy
and cosmology, as is correcting for its effects on other observables.
Mapping dust extinction over the sky is crucial to extracting correct
astrophysical quantities from observables.  In particular it is
essential in measuring the large scale structure of the universe.  For
instance, if one does not properly correct for Galactic dust extinction,
the surface number densities of galaxy behind strongly obscured regions
are reduced systematically, producing apparent void structures and
distorting the real cosmic clustering signal.

For such purposes, Schlegel, Finkbeiner \& Davis (1998; hereafter, SFD)
constructed a dust extinction map (hereafter, the SFD-map), which has
been used very extensively for a wide variety of astronomical and
cosmological studies.

The SFD-map was obtained by the following procedures: (i) making dust
temperature and emissivity maps ($0.7^{\circ}$ FWHM spatial resolution)
from COBE/DIRBE data at $100~{\mu m}$ and $240~{\mu m}$.  (ii) making a
finer resolution map for dust emission ($6.1^{\prime}$ angular
resolution) from IRAS/ISSA data at $100~{\mu m}$ using the COBE
temperature map as a calibrator.  (iii) making final maps of reddening
and extinction assuming a simple linear relation between 100$\mu m$ flux
and dust column density with a temperature correction for dust
emissivity.

For most astronomical purposes one wants an ``absorption-weighted'' dust
extinction map.  Since extinction in the SFD-map is inferred from dust
emission, however, it corresponds to an ``emission-weighted'' map.
Therefore, if the proportionality between reddening and emissivity
breaks down in some regions, for example, the SFD-map could be a poor
representation of the actual extinction map.  And in any case, there is
no guarantee that the assumptions on which the SFD-map is based are
accurate enough to construct a highly accurate and reliable account of
Galactic extinction and reddening.  For such reasons, SFD carried out
several tests of the derived maps in their initial studies.  They found
that the SFD-map reproduces intrinsic colors of elliptical galaxies as
estimated from the Mg\emissiontype{II} index (Faber et al. 1989) with an
accuracy roughly twice as good as the dust extinction map provided by
Burstein and Heiles (1978, 1982) based on HI 21-cm emission.  They also
found that the SFD-map tends to overestimate reddening at very high dust
column densities.  Also, indications of possible systematic errors
emerged from considerations of the conversion coefficient between dust
emissivity and dust reddening as determined by two different techniques:
1 - removing the correlation between the colors of brightest cluster
galaxies and the SFD-map, versus 2 - a comparison of dust emission with
galaxy number counts in the APM galaxy survey (SFD).  The coefficients
determined by these two techniques differ by approximately a factor of
two.

In this paper, we evaluate the SFD-map by a comparison with galaxy
number counts from the latest SDSS data release.  \citet{Fukugita2004}
carried out a similar study using the earlier galaxy catalog provided by
the Sloan Digital Sky Survey (SDSS:\citet{York2000}), Data Release 1
(DR1;\citet{DR1}).  They found that dust extinction values estimated
from the DR1 galaxy counts are in good agreement with the SFD ones.  The
availability of the latest catalog of SDSS (Data Release 4; \cite{DR4})
allows us to subject the SFD-map to a new and higher precision test.

%%%%%%%%%%%%%%%%%%%%%%%%%%%%%%%%%%%%%%%%%%%%%%%%%%%%%%%%
\section{The Data}
%%%%%%%%%%%%%%%%%%%%%%%%%%%%%%%%%%%%%%%%%%%%%%%%%%%%%%%%
\subsection{The Sloan Digital Sky Survey DR4}

The SDSS DR4 covers 6670 ${\rm deg}^2$ of sky area and contains 180
million objects with photometry in five pass bands, $u$, $g$, $r$, $i$,
and $z$ \citep{Fukugita1996,Gunn1998,Hogg2001,Pier2003,Blanton2003,
Ivezic2004,Smith2002,Gunn2006,Tucker2006}.  In the database, each object is
catalogued with not only its photometric properties but also the dust
extinction estimated from the SFD-map, $A_{x,{\rm SFD}}$,
($x$=$u$,$g$,$r$,$i$ and $z$).  The conversion factor from reddening,
$A_{x,{\rm SFD}}/E(B-V)$, in the SFD-map to $A_{x,{\rm SFD}}$ is
provided by SFD (see Table 6 of SFD).  To compute these factors, it was
assumed that all objects have the spectral energy distribution of an
elliptical galaxy and that Galactic dust properties are the same in all
directions (as seen from the Earth); in particular, we then adopt the
extinction curve parameter:
%%%%%%%%%%%%%%%%%%%%%%%%%%%%%%%%%
\begin{equation}
R_V \equiv A_V/E(B-V) = 3.1,
\end{equation}
%%%%%%%%%%%%%%%%%%%%%%%%%%%%%%%%%
(see \cite{EDR} and Appendix B of SFD).

\begin{figure*}[hbtp]
\vspace*{1cm}
  \begin{center}
    \FigureFile(160mm,100mm){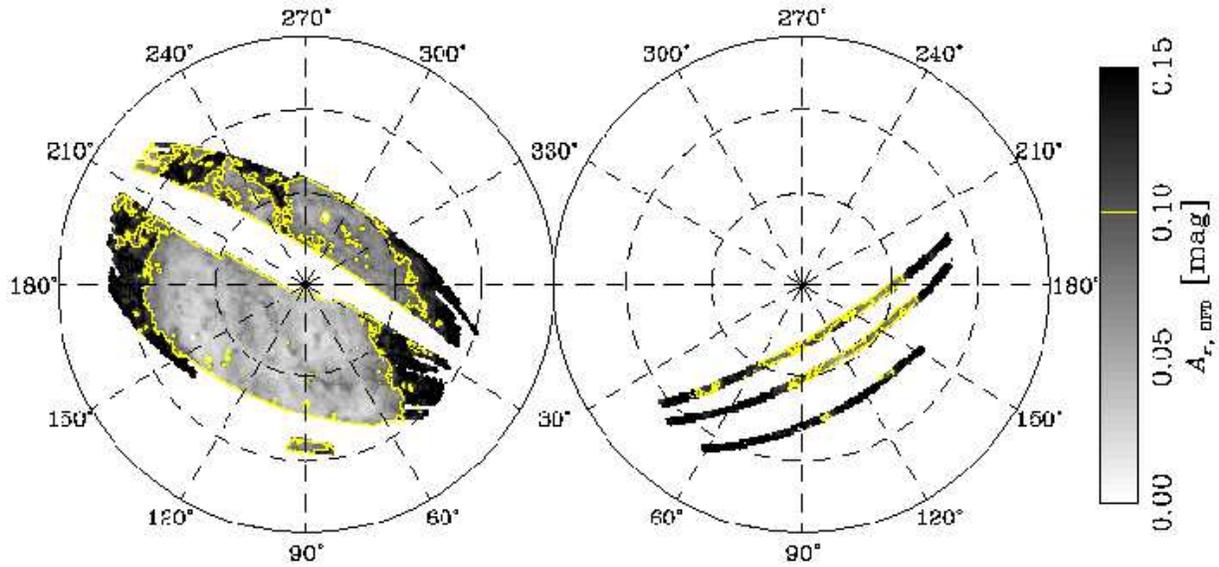}
  \end{center}
  \caption{ Photometric survey area of the SDSS DR4 in Galactic
		coordinates.  The gray scale indicates the magnitude of
		$A_{r, {\rm SFD}}$, as indicated by at the right.  The
		region in which $A_{r, {\rm SFD}} < 0.1$ is indicated by
		a contour line.  } \label{fig:survey_area}
\end{figure*}

\begin{figure}[ht]
\vspace*{1cm}
  \begin{center}
    \FigureFile(80mm,80mm){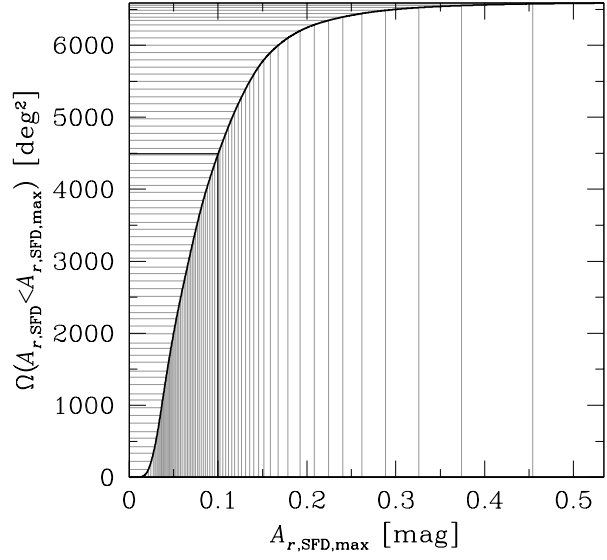}
  \end{center}
  \caption{ Cumulative distribution of sky area as a function of
		$A_{r,{\rm SFD,max}}$.  Note that $A_{r,{\rm SFD}}$ is
		less than 0.1 mag for the majority of the survey region,
		approximately 68\% in fact, as denoted by the heavy
		vertical and horizontal lines.  The vertical gray lines
		show the division of the SDSS survey region (see section
		3.1) according to extinction values.  The horizontal
		gray lines indicate the corresponding area.  }
		\label{fig:arearatio}
\end{figure}

Figure \ref{fig:survey_area} shows the SDSS DR4 photometric survey area,
and $r$-band SFD-extinction $A_{r ,{\rm SFD}}$ in the same part of the
sky.  Figure \ref{fig:arearatio} shows the cumulative area of the
regions where the value of $A_{r, {\rm SFD}}$ is lower than a certain
value $A_{r, {\rm SFD, max}}$.  We exclude the regions with relatively
high extinction, $A_{r, {\rm SFD}} > 0.534$, whose total fractional area
is less than $0.2\%$.  The gray thin lines indicate the boundary of
subregions each containing about 100 ${\rm deg}^2$ of sky (see section
3.1).  The region corresponding to $A_{r, {\rm SFD}} < 0.1$ is indicated
in both figure \ref{fig:survey_area} and figure \ref{fig:arearatio};
this is the regime in which an anomalous correlation of the SDSS galaxy
counts with the SFD-map is demonstrated in section 3.

\subsection{Constructing a Photometric Galaxy Sample}

In this paper, we initially assume that after correction for dust
extinction, the surface number density of galaxies will be homogeneously
averaged over sufficiently large areas of the sky, and that any
remaining inhomogeneities indicate an error in the estimated extinction.

For this analysis, accurate star-galaxy separation is important since
the spatial distribution of stars is not homogeneous and is likely to be
correlated with the dust distribution.  We therefore carefully
constructed a reliable photometric galaxy sample for our analysis as
follows:
%%%%%%%%%%%%%%%%%%%%%%%%%%%%%%%%%%%%%%%
\begin{enumerate}
\item False objects are discarded using photometric processing flags.
\item Masked regions are excluded.
\item A magnitude range is determined so as to ensure the reliability of the star-galaxy separation procedure.
\end{enumerate}
%%%%%%%%%%%%%%%%%%%%%%%%%%%%%%%%%%%%%%%
Details of these three steps are described below. 

\subsubsection{False objects}
\medskip

First we remove those photometric objects in the database table that
have saturated fluxes, were observed during bad sky conditions, or are
fast-moving (and thus suspected of being in the Solar System).

While one could also remove objects with interpolated fluxes, we have
chosen to retain them because they are preferentially associated with
specific bad CCD pixels and, therefore, are not randomly distributed on
the sky.  In other words, the additional photometric uncertainty
associated with pixel defects seems less likely to produce systematic
errors in our analysis than excluding a non-random spatial distribution
of objects entirely.

\subsubsection{Masks}
\medskip

The SDSS database defines masked regions on the basis of five different
conditions.  Our analysis excludes regions labeled ``BLEEDING'',
``BRIGHT\_STAR'', ``TRAIL'', or ``HOLE'', but keeps those labeled
``SEEING'' which occupy a significant fraction of the entire survey
area.  We ignore the ``SEEING'' information because the effect of
relatively bad seeing is not serious for photometry of bright galaxies.
The total area of the masked regions we exclude is about $70$ ${\rm
deg^2}$, roughly 1\% of the overall survey region.

\subsubsection{Magnitude range}
\medskip

\begin{figure*}[hbt]
  \begin{center}
    \FigureFile(170mm,120mm){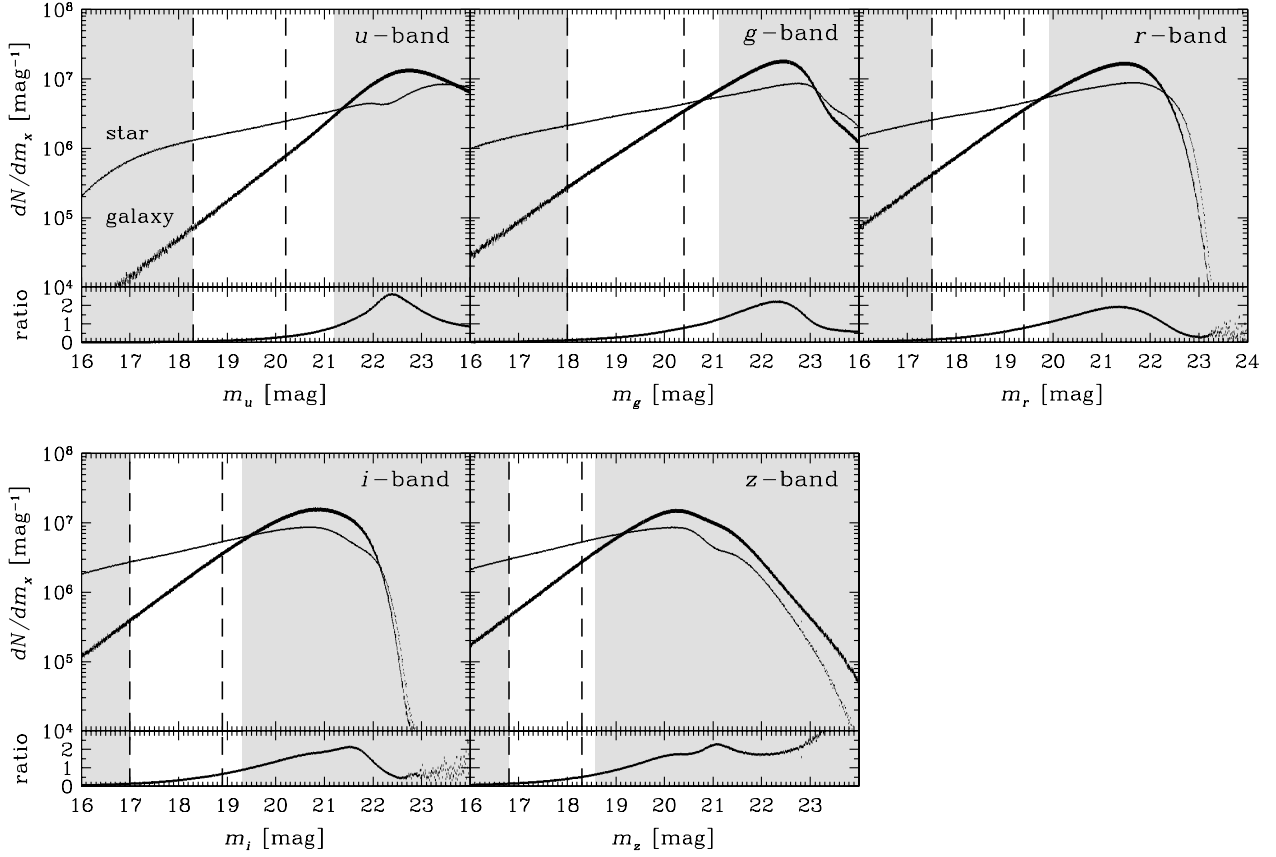}
  \end{center}
  \caption{ Differential number counts of galaxies (thick line) and
	stars (thin line) ({\it upper part of each panel}), and the
	number ratio of galaxies to stars ({\it lower part of each
	panel}) as a function of magnitude (no extinction correction).
	The two vertical dashed lines show the magnitude range in which
	we compute the surface number density of galaxies.  The upper
	left, the upper middle, the upper right, the lower left and the
	lower right panels correspond to data for $u$-, $g$-, $r$-, $i-$
	and $z$-bands, respectively.  In the $r$-band, we choose the
	magnitude range $17.5<m_r<19.4$ (the dashed vertical lines in
	the {\it upper right panel}). Even for the extinction corrected
	magnitude $m_{r,{\rm ec}}$, we choose the same magnitude range
	which is plotted as a white region. The range of analysis
	differs from band to band, and they are similarly indicated
	in the other panels.  \label{fig:histogram_galaxies_and_stars}}
\end{figure*}

We select galaxies from those DR4 objects in which the ``type''
attribute is equal to GALAXY.  The resulting number counts of galaxies
are plotted in figure \ref{fig:histogram_galaxies_and_stars} as a
function of magnitudes {\it uncorrected} for the SFD-extinction, $m_x$.

For our current purpose, we would like to construct subsets of the 
galaxy sample 
which are not contaminated by mis-identified stars to the extent 
possible. For this purpose we restrict the range of magnitudes 
in the analysis below.

Star-galaxy separation is based on the difference between the composite
model magnitude and the PSF magnitude \citep{DR2}.  The reliability of
this separation procedure depends on the magnitude of objects
\citep{Yasuda2001,Scranton2002,Strauss2002}.  In the $r$-band, the
procedure is known to be reliable down to $\sim 21$ mag.  The saturation
of stellar images typically occurs for $m_r < 15$.  To be conservative,
therefore, we chose the same magnitude range $17.5<m_r<19.4$ and
$17.5<m_{r,{\rm ec}}<19.4$ for analysis, independent of the extinction
correction; the dashed vertical lines in figure
\ref{fig:histogram_galaxies_and_stars} indicate the above range for
$m_r$, while the white regions correspond to that for $m_{r,{\rm ec}}$,
the extinction corrected magnitude.

The corresponding magnitude ranges in the other bands are similarly
indicated in figure \ref{fig:histogram_galaxies_and_stars}.  In addition
we checked that shifting the adopted magnitude ranges by up to 1
magnitude does not significantly affect our results.

The number of galaxies in the selected magnitude range (between the
dashed lines) is on the order of $10^{5}$ for the $u$-band and $10^{6}$
for the other bands.

%%%%%%%%%%%%%%%%%%%%%%%%%%%%%%%%%%%%%%%%%%%%%%%%%%%%%%%% 
\section{Analysis}

\subsection{Surface number density and average color of galaxies}

We divide the entire selected survey region into 69 subregions grouped
by their values of $A_{r,{\rm SFD}}$.  Each subregion consists of
spatially separated (disjoint) small regions of the sky with $A_{r,{\rm
SFD}}$ values in a given interval.  We define the intervals of
$A_{r,{\rm SFD}}$ such that the area of each subregion is approximately
equal ($\sim 100$ deg{$^2$}). The adopted intervals are shown by thin
vertical lines in figure \ref{fig:arearatio}.

\begin{figure*}[hbt]
 \begin{center}
  \FigureFile(170mm,120mm){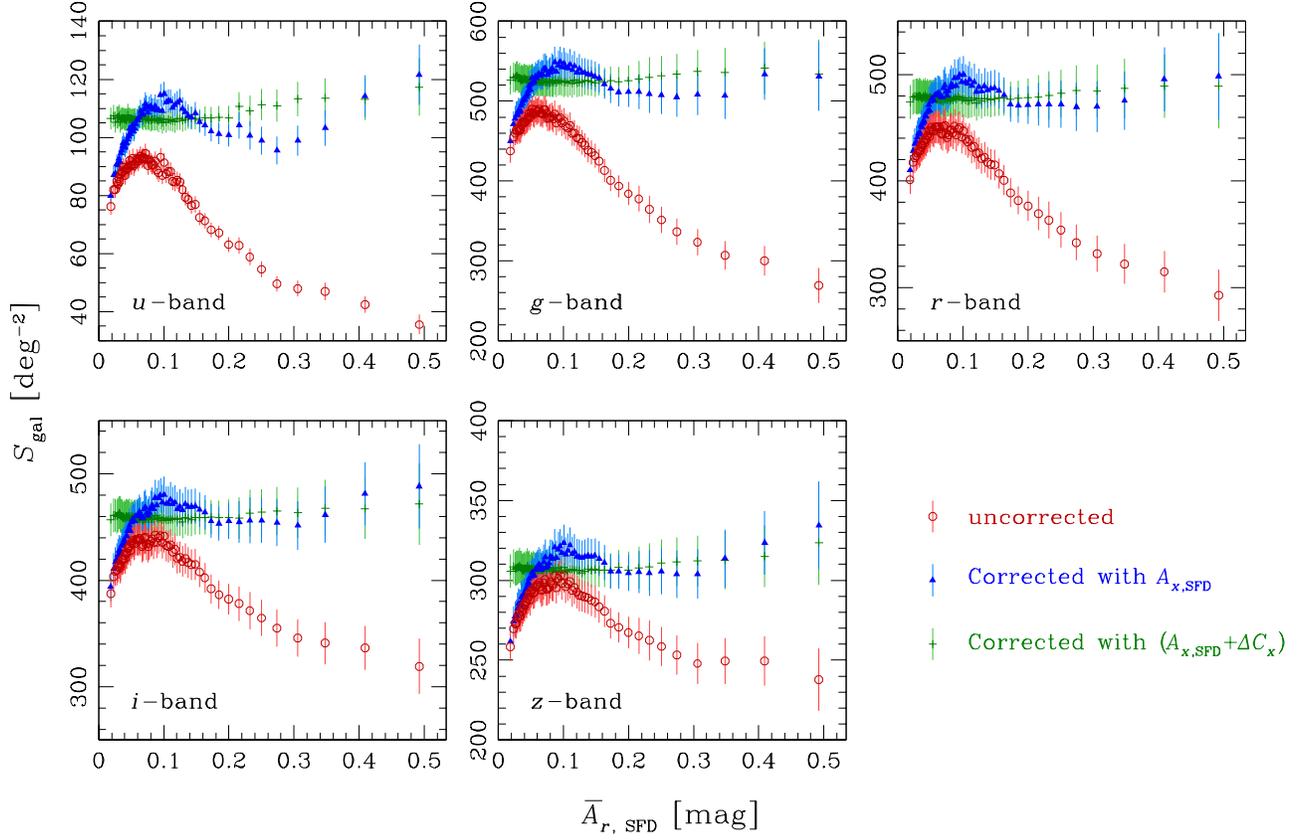}
 \end{center}
 \caption{ Surface number density of SDSS DR4 photometric sample
		galaxies in each subregion.  The horizontal axis is the
		mean SFD-extinction for the subregion, $\bar{A}_{r, \rm
		SFD}$.  The open circles (filled triangles) indicate
		that the magnitude is corrected (not corrected) using
		$A_{x,{\rm SFD}}$. The crosses indicate the magnitude
		after an additional extinction derived from the galaxy
		counts (see section 3.2).  The error bars are calculated
		using eq. \ref{eq:deltaS}.  } \label{fig:surfacedensity}
\end{figure*}

Figure \ref{fig:surfacedensity} shows the surface number densities of
galaxies, $S_{\rm gal}$, for the 69 subregions as a function of its mean
extinction value, $\bar{A}_{r,{\rm SFD}}$.  We define $\bar{A}_{r,{\rm
SFD}}$ simply by averaging ${A}_{r,{\rm SFD}}$ for all galaxies located
in the subregion.  We select the $r$-band to represent the extinction
simply because it is the central SDSS band and used for many selection
cuts, and it is indeed trivial to translate the extinction to other
bands using Table 6 of SFD.

The open circles indicate $S_{\rm gal}$ uncorrected for extinction,
while the filled triangles indicate the results after the extinction
correction.  The crosses show the results using the extinction
correction obtained from the galaxy number counts as explained in
Section 3.2; all dependence of counts on extinction is removed by
construction for the points plotted as crosses.

We compute an error estimate, $\Delta S$, for $S_{\rm gal}$ in each
subregion according to
%%%%%%%%%%%%%%%%%%%%%%%%%%%%%%%%%%%%%%%%%%
\begin{equation}
(\Delta S)^2 = {N\over \Omega^2} 
+ {{N}^2\over \Omega^3}\int_\Omega d\Omega'  w(\theta'),
\label{eq:deltaS}
\end{equation}
%%%%%%%%%%%%%%%%%%%%%%%%%%%%%%%%%%%%%%%%%%
where $N$ denotes the number of galaxies in the subregion with area
$\Omega$, and $w(\theta)$ is the angular correlation function of
galaxies.

We adopt a double power-law model for $w(\theta)$:
%%%%%%%%%%%%%%%%%%%%%%%%%%%%%%%%%%%%%%%%%%
\begin{equation}
w(\theta) = \cases{ 
	0.008 (\theta/{\rm deg})^{-0.75} & $(\theta \le 1{\rm deg})$ \cr
	0.008 (\theta/{\rm deg})^{-2.1} & $(\theta > 1{\rm deg})$ \cr
},
\end{equation}
%%%%%%%%%%%%%%%%%%%%%%%%%%%%%%%%%%%%%%%%%%
\citep{Scranton2002,Fukugita2004}.

Strictly speaking, the integral in equation (\ref{eq:deltaS}) should be
performed over the complex and disjoint shape of each subregion.
However, for simplicity we adopt a circular approximation, integrating
over $0<\theta<\theta_c$ for the actual area of the subregion, i.e., out
to $2\pi(1-\cos\theta)=\Omega$.  This approximation may slightly
overestimate the true error, but it does not affect our conclusion.  For
the typical values of $N\sim 5\times 10^4$ and $\Omega\sim 100 \deg^2$,
the second term is larger by two orders of magnitude than the first
term.  The above error-bars are plotted in figure
\ref{fig:surfacedensity}.

Figure \ref{fig:surfacedensity} indicates that the galaxy counts,
uncorrected for extinction (open circles), decrease with increasing
$\bar{A}_{r,{\rm SFD}}$ for $\bar{A}_{r,{\rm SFD}} > 0.1$, the expected
effect of Galactic dust.  However, the increase in galaxy counts with
increasing extinction for $\bar{A}_{r,{\rm SFD}} < 0.1$ is a surprise
and is the opposite of the effect expected from Galactic dust.
Furthermore, this puzzling feature remains even after the SFD extinction
correction is applied (filled triangles).  The SFD extinction correction
does properly remove the expected anti-correlation of counts and
extinction for $\bar{A}_{r,{\rm SFD}} > 0.1$.

\begin{figure*}[tbhp]
\begin{minipage}[c]{7.5cm}
  \begin{center}
    \FigureFile(70mm,70mm){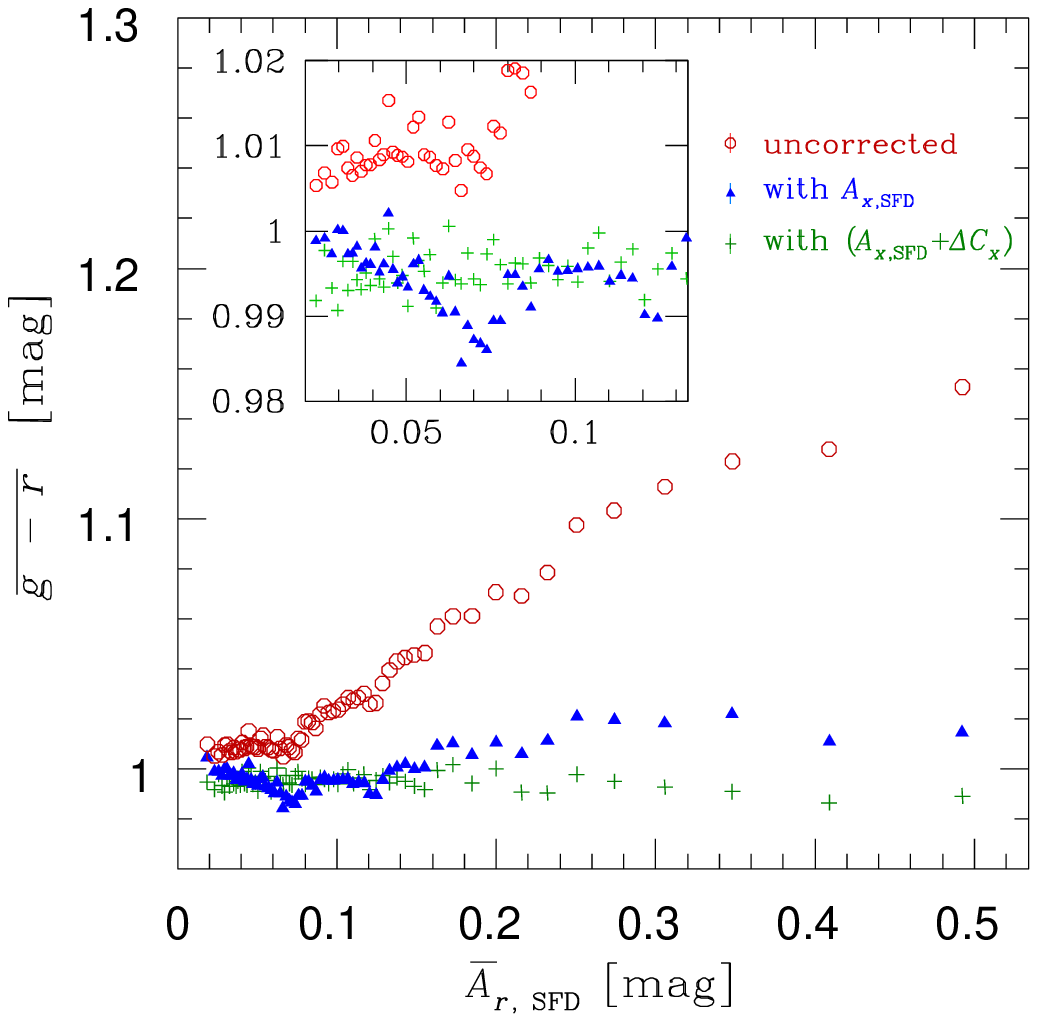}
  \end{center}
  \caption{The average $g-r$ color of galaxies in each subregion as a
	function of $A_{r,{\rm SFD}}$.  The symbols are also the same as
	in figure \ref{fig:surfacedensity}.  } \label{fig:meanGR}
\end{minipage}
%\end{figure}
%%%%%%%%%%%%%%%%%%%%%%%%%%%%%%%%%%%%%%%%%%%%%%
% Figure 5 END
%%%%%%%%%%%%%%%%%%%%%%%%%%%%%%%%
%%%%%%%%%%%%%%%%%%%%%%%%%%%%%%%% 
% Figure 6
%%%%%%%%%%%%%%%%%%%%%%%%%%%%%%%%%%%%%%%%%%%%%%%%
\hspace*{1cm}
\begin{minipage}[c]{7.5cm}
%\begin{figure}[htb]
  \begin{center}
    \FigureFile(70mm,70mm){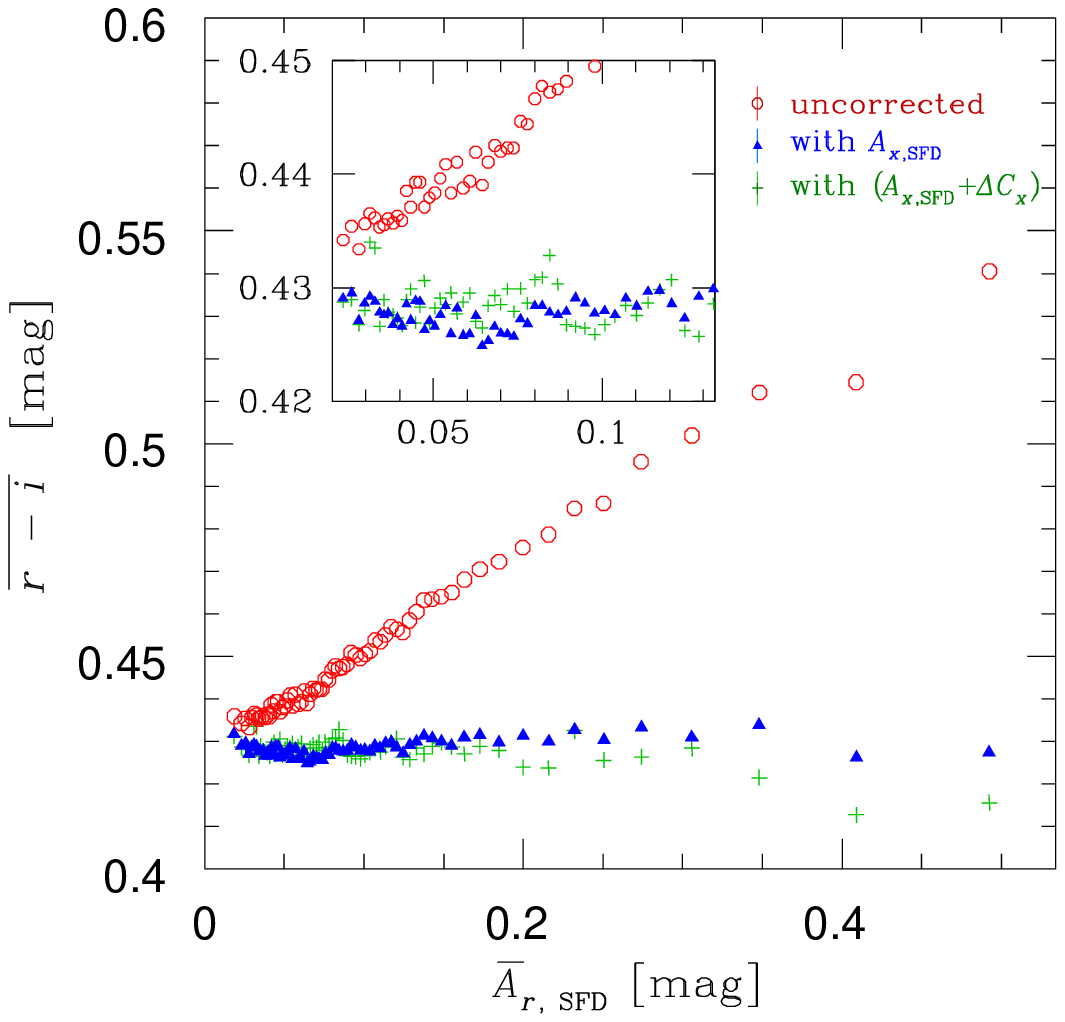}
  \end{center}
  \caption{Same as figure \ref{fig:meanGR} but for average $r-i$ color.}
  \label{fig:meanRI}
\end{minipage}
%\end{figure*}
%%%%%%%%%%%%%%%%%%%%%%%%%%%%%%%%%%%%%%%%%%%%%%%%%%
% Figure 6 END
%%%%%%%%%%%%%%%%%%%%%%%%%%%%%%%%%%%%%%%%%%%%%%%%%%%%%%%%
%%%%%%%%%%%%%%%%%%%%%%%%%%%%%%%%%%%%%%%%%%%%%%%%%%%%%%%%
% Figure 7
%%%%%%%%%%%%%%%%%%%%%%%%%%%%%%%%%%%%%%%%%%%%%%%%%%
%\begin{figure*}[hbt]
  \begin{center}
    \FigureFile(150mm,100mm){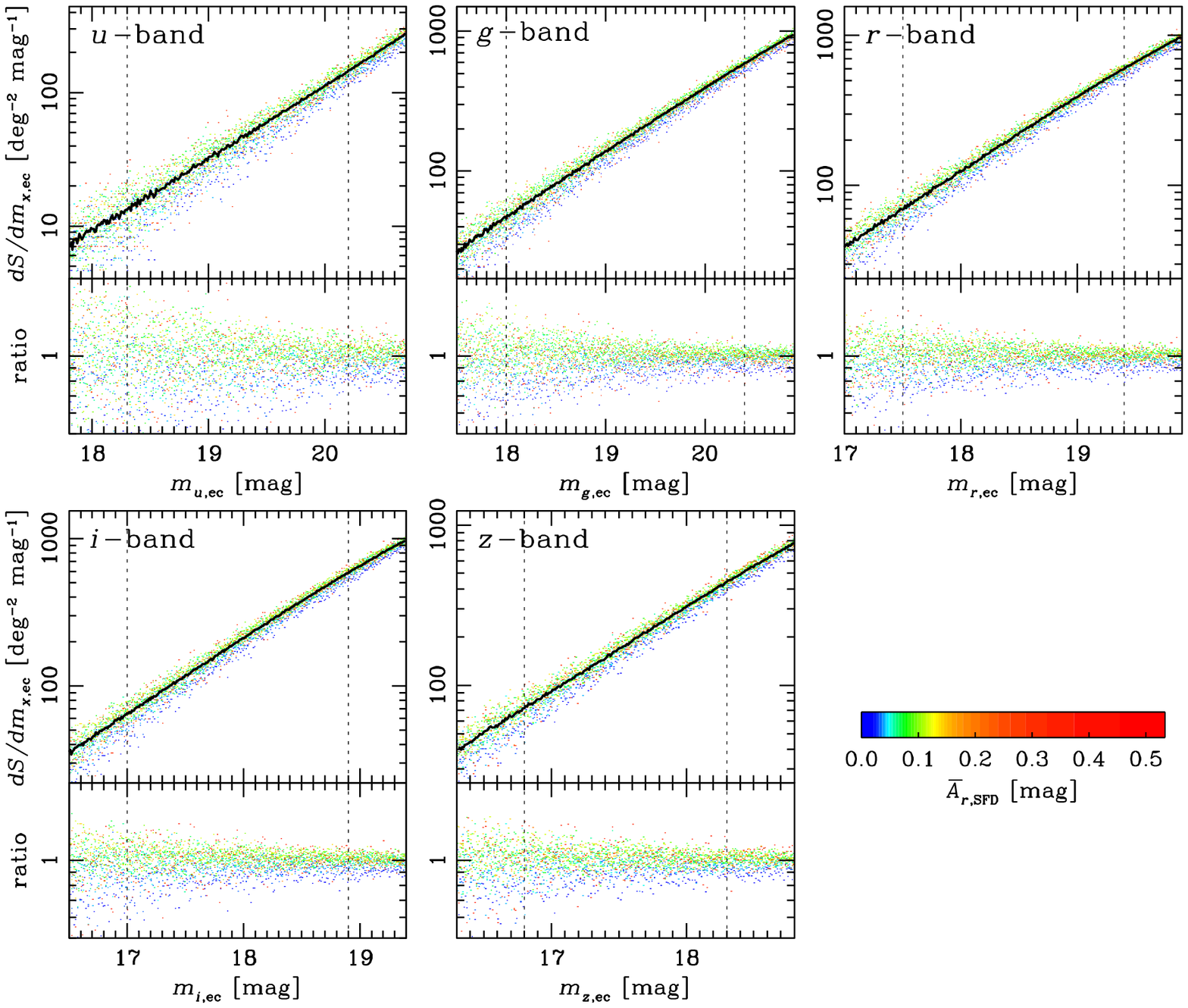}
  \end{center}
  \caption{ Differential surface number density of galaxies, $dS/dm_{x,
 {\rm ec}}$ for subregions (colored dots) and $dS_{\rm e}/m_{x,{\rm
 ec}}$ (black dots) ({\it upper part of each panel}), and the ratio of
 $dS/dm_{x, {\rm ec}}$ to $dS_{\rm e}/dm_{x,{\rm ec}}$ ({\it lower part
 of each panel}).  The bluer dots correspond to data for low
 $\bar{A}_{\rm SFD}$ regions and redder lines correspond to those for
 high $\bar{A}_{\rm SFD}$ regions.  The two vertical dashed lines are
 the same as those in figure \ref{fig:histogram_galaxies_and_stars}.
 The observational band for each panel is the same as in figure
 \ref{fig:histogram_galaxies_and_stars}.  } \label{fig:lf}
\end{figure*}

A presumably related anomaly can be also seen in the average color of
galaxies against $\bar{A}_{r,{\rm SFD}}$ for low extinctions.  Figure
\ref{fig:meanGR} shows the average $g-r$ color of galaxies corresponding
to figure \ref{fig:surfacedensity}.  The color of galaxies before the
extinction correction is redder at higher extinction for
$\bar{A}_{r,{\rm SFD}} > 0.1$, but is constant for $\bar{A}_{r,{\rm
SFD}} < 0.1$.  After the SFD correction, the galaxy color is independent
of $\bar{A}_{r,{\rm SFD}}$ for $\bar{A}_{r,{\rm SFD}} > 0.1$ as
expected.  Nevertheless the anti-correlation between $g-r$ and
$\bar{A}_{r,{\rm SFD}}$ is recognizable for $\bar{A}_{r,{\rm SFD}} <
0.1$ (see inset in figure \ref{fig:meanGR}).  A similar plot for $r-i$
color is shown in figure \ref{fig:meanRI}, but the above features are
weak, though perhaps not entirely absent.

\subsection{An additional extinction $\Delta C_x$ 
based on galaxy number counts}

In order to investigate possible reasons for the anomalous behavior of
galaxy counts at low SFD extinction, we derive an additional extinction
correction $\Delta C_x$, explicitly constructed to make the two
quantities independent.  In other words, by using the fact that $S_{\rm
gal}$ should be independent of $A_{r,{\rm SFD}}$ if the galaxy
magnitudes are properly corrected for Galactic dust extinction, we solve
for an additional extinction $\Delta C_x$ relative to $A_{x,{\rm SFD}}$
that enforces this behavior.

Figure \ref{fig:lf} shows the differential surface number density of
galaxies , $dS/dm_{x,{\rm ec}}$. Note that the magnitudes, $m_{x,{\rm
ec}}$, in this figure refer to the values corrected for the extinction
using $A_{x,{\rm SFD}}$.  The black lines are the differential surface
number density of galaxies in the entire survey region, $dS_{\rm
e}/dm_{x,{\rm ec}})$.  Those for each subregion, $dS/dm_{x,{\rm ec}}$,
are plotted in colored dots according to the value of $\bar{A}_{r,{\rm
SFD}}$ indicated in the color bar.

It is clear that low $A_{r,{\rm SFD}}$ subregions (blue dots)
preferentially lie below $dS_{\rm e}/dm_{x,{\rm ec}}$, while redder ones
lie above it.  This systematic trend is simply another representation of
the anomalies shown in figure \ref{fig:surfacedensity}.  We now find the
best fit values for the additional correction $\Delta
C_x(\bar{A}_{x,{\rm SFD}})$ by a $\chi^2$ analysis for the shifted
number density $dS/dm_{x,{\rm ec}}-\Delta C_x)$ in each subregion, and
$dS_{\rm e}/dm_{x,{\rm ec}}$.

The results for $\Delta C_x$ as a function of $\bar{A}_{x,{\rm SFD}}$ in
the five different bands are plotted in figure
\ref{fig:delta_extinction}.  All the panels show the same systematic
behavior for $A_{r,{\rm SFD}} < 0.1$ which is required to cancel the
anomaly in figure \ref{fig:surfacedensity} (crosses).  The $S_{\rm gal}$
with the additional correction is indeed completely independent of
$\bar{A}_{r,{\rm SFD}}$ within the quoted error-bars, which simply
confirms that the calculation has been carried out correctly.

The same correction simultaneously removes the anomaly in the $g-r$
color of galaxies in figure \ref{fig:meanGR} (crosses).  This, however,
is not a circular result and thus lends some credibility to the exercise
since the same hypothetical additional dust component required to remove
the anomaly in the counts need not also be one which eliminates the
color anomaly.

Despite being internally self-consistent and simultaneously satisfying
both count and color correction constraints with a single free function,
the physical implications of this explanation of the anomalies,
basically that they are due to an unknown component of Galactic dust,
are not particularly plausible.  In particular, the amplitudes of the
corrections are surprisingly large, of order 0.1 to 0.2 magnitudes, and
rather insensitive to the band in which they are defined (about half as
large in $z$-band as in $u$-band).  In the next section we subject the
unknown dust component explanation to further tests and suggest an
alternative and more plausible possibility.

\section{Interpretation}

In this section we will consider possible explanations for the
surprising anomaly reported in the previous section, namely that there
are fewer SDSS galaxies (in both raw and corrected counts) in regions of
the sky where the SFD map indicates the least extinction, precisely the
opposite of the trend one would naively expect.

\subsection{Effects on counts of more distant objects}

If the anomalous excess counts of SDSS galaxies at low values of
$A_{x,{\rm SFD}}$ reported above were actually due to systematic
underestimation of the extinction at low extinction in the SFD map ({\it
i.e.}, the most direct interpretation), then the same effect would also
be apparent in the distribution of more distant cosmic objects.  Here we
show that this is not the case.

\begin{figure}[hbt]
  \begin{center}
    \FigureFile(80mm,80mm){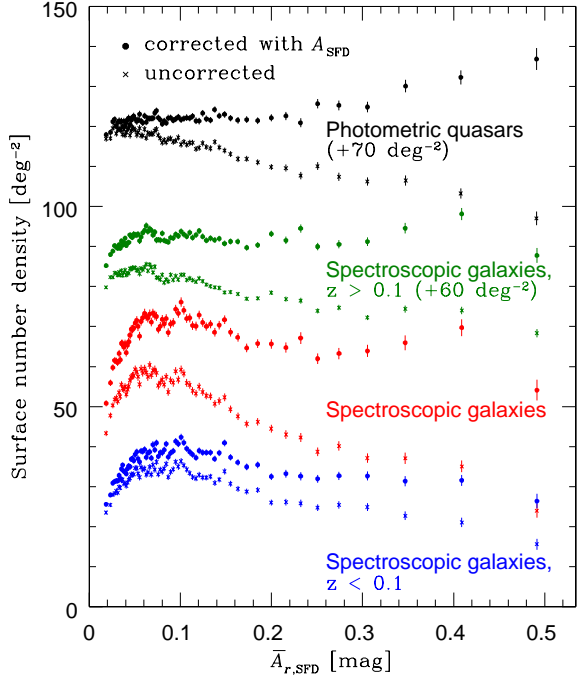}
  \end{center}
  \caption{Surface number density of SDSS photometric quasars and
	spectroscopic galaxies as a function of $\bar{A}_{r, {\rm
	SFD}}$.  The filled circles and the crosses indicate the results
	with and without an extinction correction using $A_{x,{\rm
	SFD}}$, respectively. Just for clarity, the data points of
	photometric quasars and spectroscopic galaxies with $z>0.1$ are
	shifted upward by $+70 {\rm deg}^{-2}$ and $+60 {\rm deg}^{-2}$,
	respectively.

 The error bars are 1-$\sigma$ Poisson
	error.  } \label{fig:surfacenumberdensityQSO}
\end{figure}

In order to see the extent to which the anomaly in $S_{\rm gal}$ depends
on the distance of the sample, we repeat the same analysis separately
for samples of nearby and more distant SDSS spectroscopic galaxies as
well as for a SDSS photometric quasar sample \citep{Richards2004}.  The
results are summarized in figure \ref{fig:surfacenumberdensityQSO}.  The
different symbol colors indicate the surface number densities of
photometric quasars (blue) and spectroscopic galaxies (red).  The latter
is further divided into two subsamples according to the redshift;
$z>0.1$ (green) and $z<0.1$ (yellow).  The open circles refer to the
uncorrected data, while the filled ones to the data which are corrected
using $A_{x,{\rm SFD}}$.  Note that the selection criteria for the
photometric quasar sample already make use of the SFD-extinction
correction, but this should not invalidate our use of the data.

Clearly, the spectroscopic galaxy sample shows a similar anomaly at low
$A_{r,{\rm SFD}}$ to the one seen in the photometric galaxy sample.
However, the effect is much weaker in the $z>0.1$ subsample than in the
$z<0.1$ one.  Moreover, the photometric quasar sample shows little, if
any, sign of the effect at low extinction.

We find no obvious correlation between the ${\bar A}_{r,{\rm SFD}}$ and the
observed (uncorrected) mean surface density of QSO's for the cells in
the range ${\bar A}_{r,{\rm SFD}}<0.15$. The galaxy clustering anomaly acts to
shuffle cells in the ${\bar A}_{r,{\rm SFD}}<0.15$ range, producing spuriously
higher and lower values of $A_{\rm SFD}$ around the level of any
underlying galactic extinction. We expect that the anomaly is clustered
like bright galaxies and therefore it should be unrelated to the surface
density of distant QSO's. Hence, where we are dominated by this, we may
expect a flatter relation between the QSO counts and $A_{\rm SFD}$, as
observed, washing out the weak underlying galactic extinction. A flat
relation is also clear for the $z>0.1$ galaxy sample, indicating they
are in the background, relatively unrelated to the anomaly.

The fact that the SFD extinction corrected counts of distant objects are
uncorrelated with the extinction while those of nearby objects exhibit
the low $A_{x,{\rm SFD}}$ anomaly strongly suggests that the effect is
not solely a problem with the extinction map but must also be connected
to the nearby galaxies as well.  In order to further elucidate the
situation, we now turn to an alternative indicator of Galactic
extinction.

\subsection{Galaxy surface number density versus 
an {\rm H{\sc i}} extinction map}

\begin{figure}[tbh]
  \begin{center}
    \FigureFile(80mm,80mm){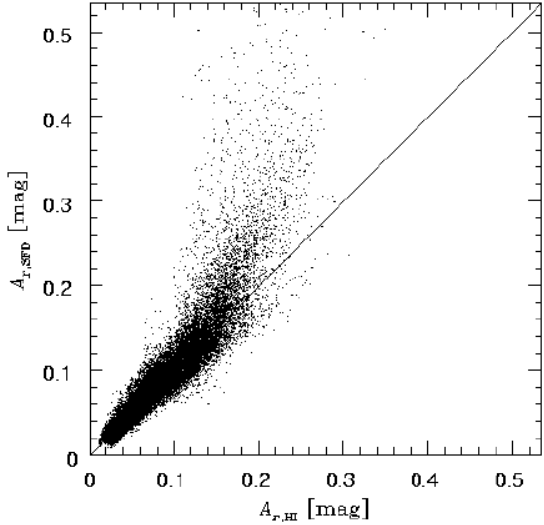}
  \end{center}
  \caption{Correlation between $A_{r,{\rm SFD}}$ and $A_{r,{\rm
 HI}}$. The latter is estimated from the Leiden-Dwingeloo {\rm H{\sc i}}
 (21 cm) map \citep{Hartmann1996}.  } \label{fig:HI_vs_SFD}
\end{figure}

\begin{figure}[tbh]
  \begin{center}
    \FigureFile(80mm,80mm){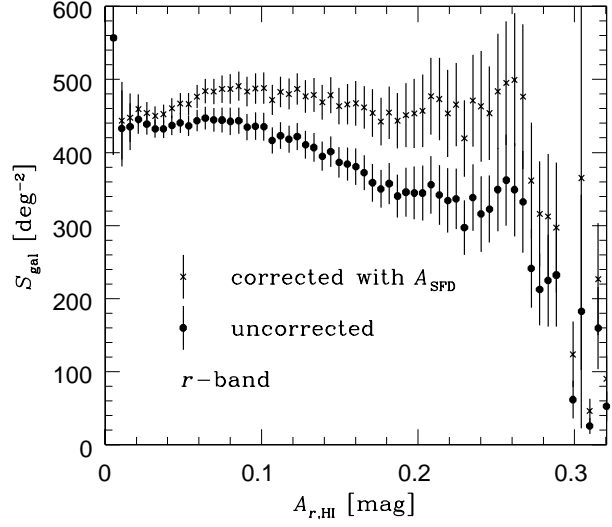}
  \end{center}
  \caption{ Surface number density of galaxies as a function of
$A_{r,{\rm HI}}$. Filled circles and crosses indicate $S_{\rm gal}$
before and after correcting for the extinction using $A_{r,{\rm SFD}}$,
respectively.  } \label{fig:HI_vs_Sgal}
\end{figure}

We next investigate the correlation between the surface number density
of SDSS photometric galaxies and the Leiden-Dwingeloo {\rm H{\sc i}} (21
cm) map \citep{Hartmann1996}.  The {\rm H{\sc i}} map can be used as a
tracer of dust column density in optically thin regions.  According to
SFD, the conversion from the {\rm H{\sc i}} map to $r$-band extinction,
$A_{r,{\rm HI}}$, is given as follows.
%%%%%%%%%%%%%%%%%%%%%%%%%%%%%%%%%%%%%%%%%%%%%
\begin{eqnarray}
& & A_{r,{\rm HI}} \;\;[{\rm mag}]  
=  2.751 \times  0.0184 \;\;[{\rm mag/(MJy/sr)}] \cr
& & \hspace{1cm} \times ~0.01222 \;\;[{\rm (MJy/sr)/(K\;km/s)}] \cr
& & \hspace{1cm} \times  
\sum_{-72 < v_{\rm LSR} \;[{\rm km/s}]< +25} 
1.03T_{a}(v_{\rm LSR})\;\;[{\rm K\;km/s}],
\end{eqnarray}
%%%%%%%%%%%%%%%%%%%%%%%%%%%%%%%%%%%%%%%%%%%%%
where $T_a$ is antenna temperature in the {\rm H{\sc i}} map. The 21 cm
emission is summed up in the velocity range ${-72 < v_{\rm LSR} < +25}$.
{\it The velocity cut is very important because it rigorously excludes
extragalactic effects on the Galactic extinction map.} In addition, it
also removes the contribution of high velocity clouds which presumably
have very little dust. Figure \ref{fig:HI_vs_SFD} shows the correlation
between $A_{r,{\rm HI}}$ and $A_{r,{\rm SFD}}$.  In the low extinction
region ($A_{r, {\rm SFD}} <0.1$), the correlation is almost linear.
Figure \ref{fig:HI_vs_Sgal} shows the surface number density (no
extinction correction) as a function of $A_{r,{\rm HI}}$. In this
figure, there is very little, if any, anomaly of the type seen in figure
\ref{fig:surfacedensity}.

\subsection{Likely Contamination of the SFD Extinction Map 
by Extragalactic FIR Emission}

In attempting to understand the low extinction anomaly it is useful to
keep in mind that $A_{x,{\rm SFD}}$ is basically proportional to the
100$\mu m$ flux observed in the IRAS/ISSA sky map, aside from modest
dust temperature corrections derived from the COBE/DIRBE sky map at 100
and 240 $\mu m$.  This means that the positive correlation between SDSS
galaxy counts and $A_{x,{\rm SFD}}$ can also be thought of as being a
positive correlation with the 100$\mu m$ flux.  This suggests the
obvious and plausible hypothesis that the correlation is simply due to
the 100$\mu m$ emission of the galaxies themselves, both those directly
detected in the Sloan survey as well as other (optically) fainter ones
which follow the galaxy clustering structures traced by the SDSS
galaxies.

\begin{figure}[tbh]
  \begin{center}
    \FigureFile(80mm,80mm){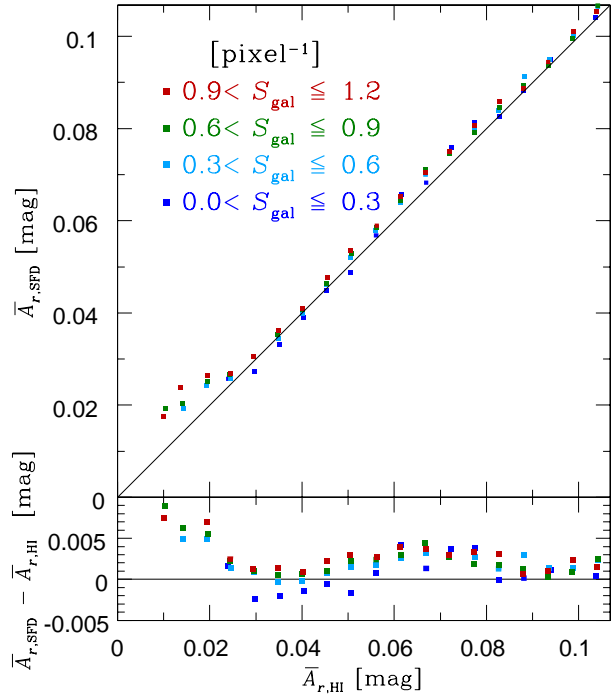}
  \end{center}
  \caption{Correlation between $A_{r,{\rm SFD}}$ and $A_{r,{\rm
HI}}$ binned according to $S_{\rm gal}$.  } \label{fig:12}
\end{figure}

Figure \ref{fig:12} combines the information in Figures
\ref{fig:HI_vs_SFD} and \ref{fig:HI_vs_Sgal} and illustrates that the
tight correlation of $\bar A_{r, {\rm SFD}}$ and $\bar A_{r,{\rm HI}}$
extends to very small extinctions. More importantly, we note that 
${\bar A}_{r,
{\rm SFD}}$ is slightly but systematically larger for larger $S_{\rm
gal}$, which is clearly exhibited in the lower panel.  On the other
hand, the weak departure from linearity below 0.02 is an artifact due to
the averaging procedure over bins of $\bar A_{r,{\rm HI}}$ with the
presence of noises; we made sure that the data points shift to the
right, instead of downward, when the average is taken over the bins of
$\bar A_{r,{\rm SFD}}$.

This again suggests that the galaxy count behavior in the low $A_{\rm
SFD}$ regions is related to some effect in the construction of the
SFD map and not to some previously unknown component of
Galactic dust.

The hypothesis can be most clearly understood by examining the shape of
the relationship between raw galaxy counts and $A_{x,{\rm SFD}}$ shown
by the red points in figure \ref{fig:surfacedensity}.  At SFD
extinctions above the peak in ${S}_{\rm gal}$ in each band, we suppose
that the observed 100$\mu m$ flux is indeed dominated by Galactic dust
emission and thus that the inferred SFD extinction is thus a good
indicator of the actual extinction; as expected the raw galaxy counts
then fall with increasing extinction.  However, at SFD extinctions below
this peak, we suppose that extragalactic FIR emission is making a
substantial, perhaps dominant, contribution to the total 100$\mu m$ flux
and thus producing the observed positive correlation between galaxy
counts and SFD inferred extinction.  In other words, we can still
observe the intrinsic correlation of extragalactic optical and FIR
emission where the Galactic dust emission is weak enough to unveil it.

Of course, SFD were aware of the possibility of extragalactic
contributions to the observed IRAS/ISSA 100$\mu m$ flux and attempted to
minimize it by both subtracting the contributions of approximately
10,000 known point sources from the sky map and then subtracting a
uniform flux density of $\nu I_\nu \sim 25$ ${\rm nW m^{-2} sr^{-1}}$ at
100$\mu m$ in order to remove the {\it mean} contribution of fainter,
unresolved point sources.  However, it is clear that neither procedure
can remove the {\it fluctuations} in the background extragalactic FIR
light due to faint sources.  It is these fluctuations, interpreted as
variations in Galactic extinction in the SFD map, which we believe to be
responsible for the observed anomaly.

We may now investigate this hypothesis quantitatively by comparing the
extinction inferred from galaxy number counts (as in section 3.2) with
that inferred from 100$\mu m$ flux by SFD on the assumption that there
is actually a contribution to this flux from galaxies at the expected
level.

First we note that \citet{Finkbeiner2000} detected the infrared
background at a level of $\nu I_\nu \sim 25$ ${\rm nW m^{-2} sr^{-1}}$
at 100$\mu m$.  This corresponds to about 0.015 mag in $E(B-V)$, ($\sim$
0.04 mag in $A_r$).  This value is comparable to the $A_{r,{\rm SFD}}$
values in the region where the anomalous positive correlation of surface
number density of galaxies exist with $A_{r,{\rm SFD}}$ is observed.  By
itself this indicates that the order-of-magnitude strength of the
hypothesized contamination is appropriate to explain the anomaly.

Second if fluctuations in galaxy surface density on the sky are
attributed entirely to dust, the ratio of the observed surface number
density in a certain direction, $S_{\rm gal}$ and the average of the
surface number density of galaxies over the whole sky, $\bar{S}_{\rm
gal}$ can be written as
%%%%%%%%%%%%%%%%%%%%%%%%%%%%%%%%%%%%%%%%%%%%%%%%%%%%%%%%%%%%%%
\begin{equation}
\label{eq:ratio_Sgal}
S_{x,{\rm gal}}/\bar{S}_{x,{\rm gal}} = 10^{-\gamma_x\Delta C_x},
\end{equation}
%%%%%%%%%%%%%%%%%%%%%%%%%%%%%%%%%%%%%%%%%%%%%%%%%%%%%%%%%%%%%%
where $\gamma_x$ and $\Delta C_x$ are the slope of $dS/dm_x$ and the
correction to the $A_{x,{\rm SFD}}$ required to yield uniform galaxy
counts across the sky (see section 3.2), respectively.  On the other
hand, if {\it fluctuations} in galaxy surface density on the sky comes
from the contamination of total FIR flux by extragalactic flux, the
expected correction to the $A_{x,{\rm SFD}}$, $\Delta A_{x,{\rm IR}}$
can be written as
%%%%%%%%%%%%%%%%%%%%%%%%%%%%%%%%%%%%%%%%%%%%%%%%%%%%%%%%%%%%%%
\begin{equation}
\label{eq:deltaA1}
\Delta A_{x,{\rm IR}} = k_{x/r} 
{0.04\over \bar{S}_{x,{\rm gal}}} (\bar{S}_{x,{\rm gal}}-S_{x,{\rm gal}}),
\end{equation}
%%%%%%%%%%%%%%%%%%%%%%%%%%%%%%%%%%%%%%%%%%%%%%%%%%%%%%%%%%%%%%
where $k_{x/r}$ is the conversion factor from $A_r$ to $A_x$.  Therefore,
 the relation between $\Delta C_x$ and $\Delta A_{x,{\rm IR}}$ is
%%%%%%%%%%%%%%%%%%%%%%%%%%%%%%%%%%%%%%%%%%%%%%%%%%%%%%%%%%%%%%
\begin{equation}
\label{eq:deltaA2}
\Delta A_{x,{\rm IR}} = 0.04k_{x/r}(1-10^{-\gamma_x\Delta C_x}).
\label{eq:dustToIR}
\end{equation}
%%%%%%%%%%%%%%%%%%%%%%%%%%%%%%%%%%%%%%%%%%%%%%%%%%%%%%%%%%%%%%

\begin{figure*}[hbt]
  \begin{center}
    \FigureFile(170mm,120mm){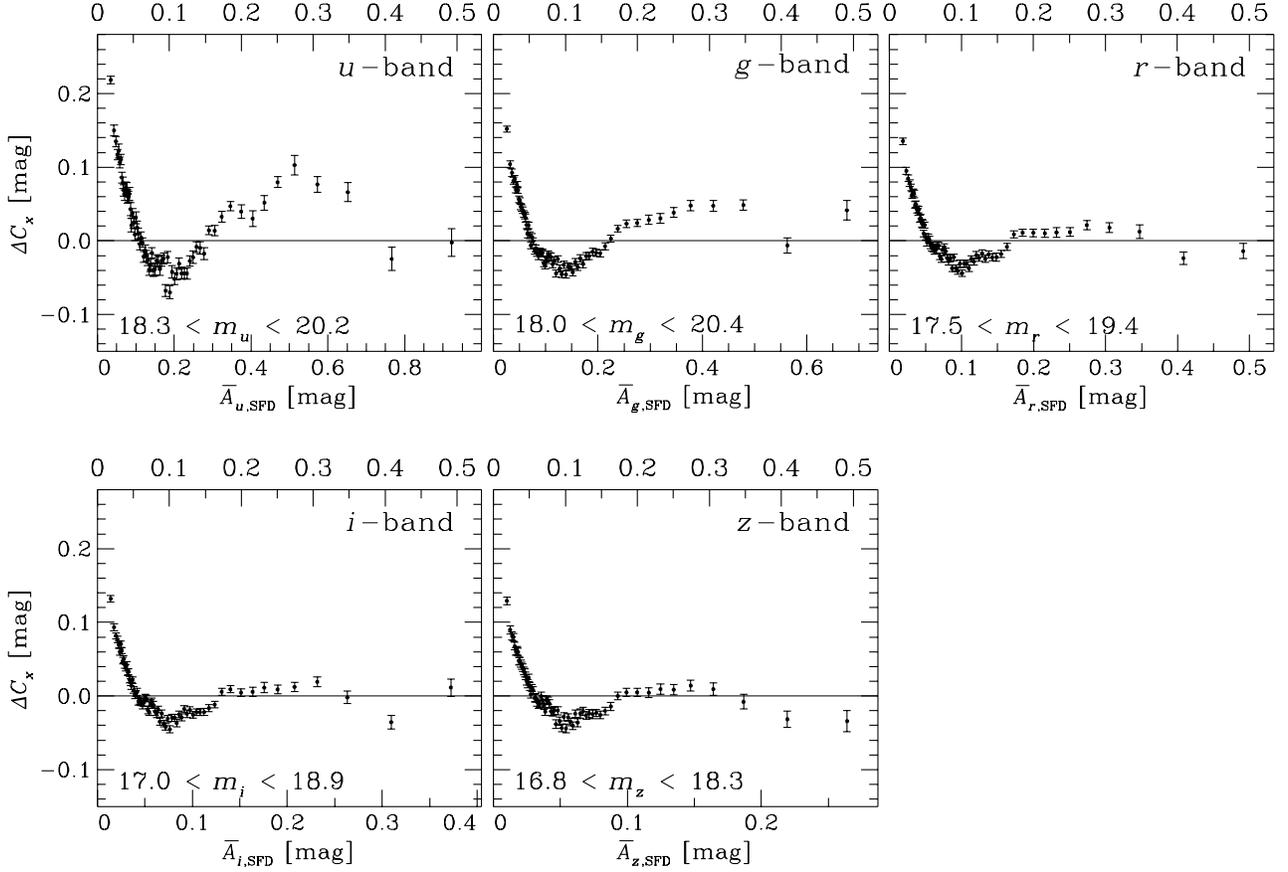}
  \end{center}
  \caption{The additional extinction $\Delta C_x$ required to give a
		constant corrected galaxy surface number density in each
		subregion.  The horizontal axis is the mean $A_{x,{\rm
		SFD}}$ in each band and is scaled so that the relative
		positions of a subregion in each of the five panels are
		the same (the upper scales indicate the corresponding
		$A_{r,{\rm SFD}}$ values).  Error bars are evaluated by
		the $\chi^2$ minimization procedure to fit
		$dS/dm_{x,{\rm ec}}-\Delta C_x)$ to $dS_{\rm
		e}/dm_{x,{\rm ec}}$.  } \label{fig:delta_extinction}
\end{figure*}

\begin{figure*}[htb]
  \begin{center}
    \FigureFile(170mm,110mm){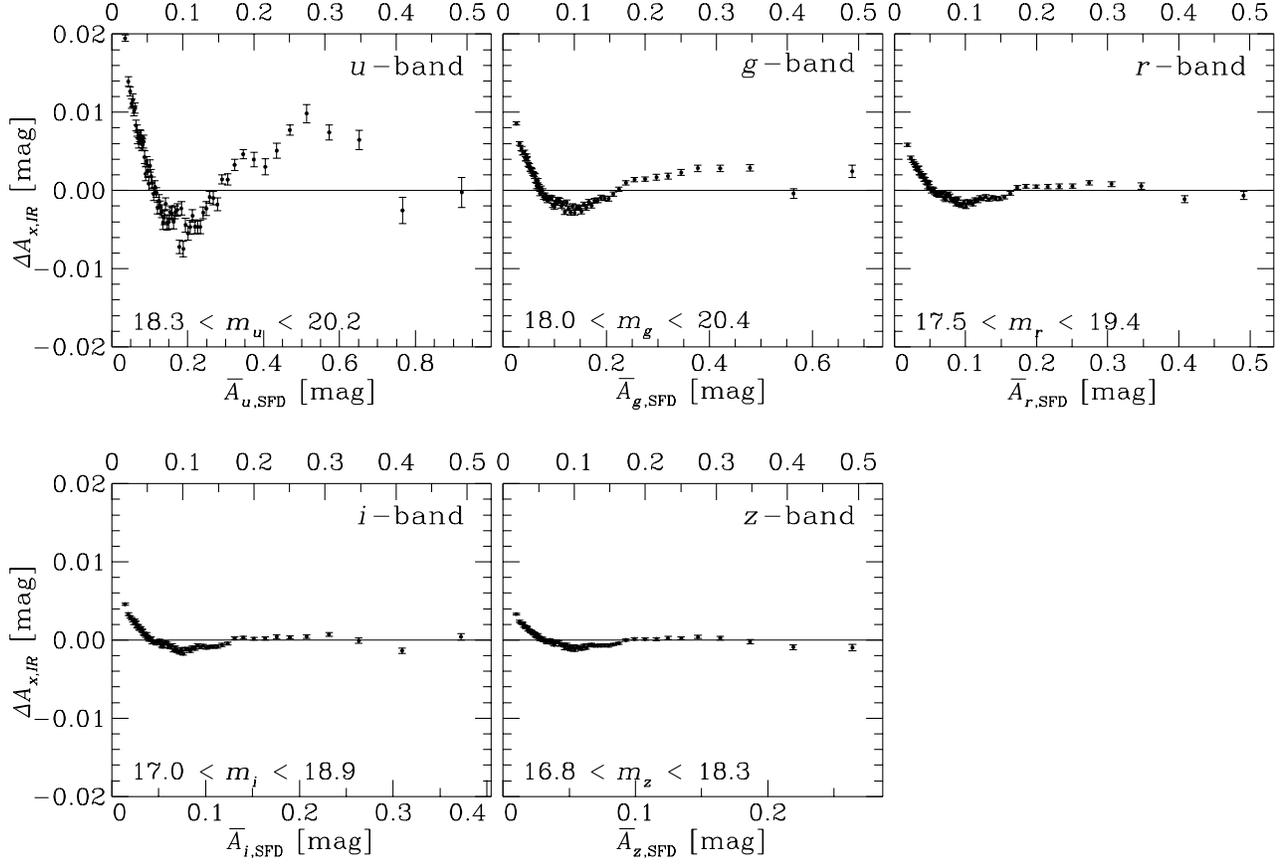}
  \end{center}
  \caption{The mean correction $\Delta A_{x,{\rm IR}}$ to $A_{x,{\rm
SFD}}$ implied by the hypothesis of extragalactic FIR contamination of
the SFD-map.  The vertical axis is the mean extinction in the SFD-map
which is actually due to this contamination rather than actual Galactic
dust (see eq.[\ref{eq:dustToIR}]).  } \label{fig:delta_A_IR}
\end{figure*}

Note that equations (\ref{eq:ratio_Sgal}) to (\ref{eq:deltaA2}) should
be interpreted as an order-of-magnitude relation, but illustrate the
qualitative effect of the IR emission of galaxies on the estimate of the
{\it Galactic} extinction.  Figure \ref{fig:delta_A_IR} shows $\Delta
A_{x,{\rm IR}}$ converted by equation (\ref{eq:dustToIR}) from $\Delta
C_x$ which is shown in figure \ref{fig:delta_extinction}. {\it Thus the
amplitude of $\Delta A_{x,{\rm IR}}$ is about one-tenth that of $\Delta
C_x$.}  This quantitative analysis supports the extragalactic FIR
contamination hypothesis.  It shows that rather small systematic errors,
of order hundredths of a magnitude, in the SFD extinction values (at low
extinction) can plausibly explain the surprisingly large values of
$\Delta C_x$ shown in figure \ref{fig:delta_extinction}.

Note that one cannot use the above relationship directly as a correction
to $A_{\rm SFD}$ because it only applies on average (over 100 ${\rm
deg}^2$ regions of the sky) but would not improve the extinction
estimate for individual pixels in the SFD map.  In principle, some
combination of the observed SDSS galaxy counts in each SFD pixel and the
observed 100$\mu m$ flux could be combined to provide an improved
estimate of the actual extinction; however, in practice this is not
likely to be very successful due to the Poisson noise in the SDSS map at the
SFD angular resolution (corresponding to only of order a single survey
galaxy per pixel).  Moreover, the absolute correction to the SFD
extinction would be quite small, of order 0.01-0.02 magnitudes, although
the relative correction could be substantial on the sight-lines with
already very low SFD extinctions.

\begin{figure}[tbh]
  \begin{center}
    \FigureFile(80mm,80mm){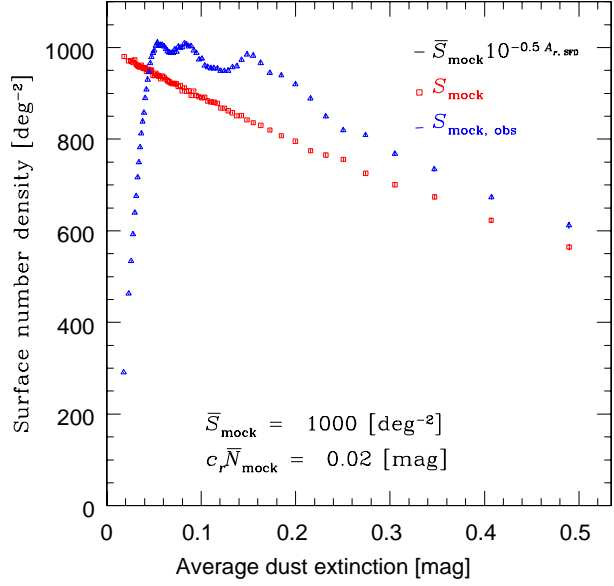}
  \end{center}
  \caption{Simulated surface number density of galaxies in a mock survey
	as a function of $\bar{A}_{r, {\rm c}}$ or $\bar{A}_{r,{\rm
	SFD}}$. The open triangle (square) corresponds to extragalactic
	FIR emission contamination (or absence thereof) of the FIR
	Galactic dust emission. We set the surface number density of
	mock galaxies ,$\bar{S}_{\rm mock}$, to 1000 ${\rm deg^{-2}}$
	and the typical contribution to the $A_{r, {\rm SFD}}$ of those
	galaxies, $c_r\bar{N}_{\rm mock}$, to 0.02 mag.  }
	\label{fig:irback}
\end{figure}

As a final test of the hypothesis, we simulate the expected effect using
mock data and show that it is consistent with the observations.  We
constructed a mock galaxy sample with the Poisson noise in the same
region of the sky as the SDSS survey area, and counted the mock galaxies
in pixels as defined in SFD map (each pixel has 5.635 ${\rm min^2}$
area).  Then we use the existing SFD-map as the ``true'' dust extinction
and add extra ``extinction'', inferred from the assumed extra 100$\mu m$
flux, proportional to the number of galaxies in each pixel according to
%%%%%%%%%%%%%%%%%%%%%%%%%%%%%%%%%%%%%%%%%%%%%%%%%%%%%%%%%%%%%%
\begin{equation}
A_{r, {\rm c}} = A_{r, {\rm SFD}} + c_r(N_{\rm mock}-\bar{N}_{\rm mock}),
\label{eq:newExtinction}
\end{equation}
%%%%%%%%%%%%%%%%%%%%%%%%%%%%%%%%%%%%%%%%%%%%%%%%%%%%%%%%%%%%%%
where $A_{r, {\rm c}}$ is the calculated $r$-band model extinction when
the extragalactic FIR flux is not distinguished from the infrared
emission from Galactic dust.  Thus $c_r$ and $N_{\rm mock}$ are the mean
conversion factor and number of mock galaxies in a pixel. We then
computed the ``observed'' surface number density of mock galaxies,
$S_{\rm mock, obs}$, as follows:
%%%%%%%%%%%%%%%%%%%%%%%%%%%%%%%%%%%%%%%%%%%%%%%%%%%%%%%%%%%%%%
\begin{equation}
S_{\rm mock, obs} = S_{\rm mock} 10^{(-0.5 A_{r, \rm c})},
\end{equation}
%%%%%%%%%%%%%%%%%%%%%%%%%%%%%%%%%%%%%%%%%%%%%%%%%%%%%%%%%%%%%%
where $S_{\rm mock}$ is the true surface number density of mock
galaxies.  The open square and open triangle in figure \ref{fig:irback}
show $S_{\rm mock}$ and $S_{\rm mock, obs}$, respectively.  It exhibits
the same qualitative behavior seen in figure \ref{fig:surfacedensity}
and is even more dramatic due to the absence of all noise and
measurement errors.

\section{Summary}

We have compared the SFD Galactic extinction map to the number counts of
SDSS photometric galaxies.  For SFD extinctions above 0.1 to 0.2
magnitudes, depending on the band, we find the two types of estimates to
be in tolerable agreement on average.  However, for smaller values of
the SFD extinction, we find a substantial and systematic disagreement In
particular, we find that that average galaxy counts (surface density on
the sky) {\it increase} and average galaxy colors become slightly {\it
redder} with {\it increasing} SFD extinction, precisely the opposite of
expected dust effects.  This low SFD extinction regime exhibiting
anomalous behavior includes approximately 68\% of the high Galactic
latitude sky covered by the SDSS, as well as most other observations of
extragalactic objects.

Although one could explain the observations with a hypothetical
component of Galactic dust which is somehow anti-correlated the 100$\mu
m$ flux based SFD-map, this does not seem physically plausible.  In
addition there is no sign of such a component in {\rm H{\sc i}} based
extinction maps.  Moreover, the surface number density of distant
quasars does not exhibit any such anomaly, as would be expected if it
arose from some unknown component of Galactic dust.  Therefore, we
conclude the effect is not due to major deficiencies in the SFD
extinction map.

An alternative and more reasonable explanation is provided by the
hypothesis that residual FIR emission from external galaxies
contaminates the signal from Galactic dust and becomes the dominant
contribution on sight-lines where the actual Galactic extinction and,
thus, dust emission are low.  We show that this explanation is
quantitatively plausible and consistent with the observed effect.  An
extragalactic FIR flux corresponding to only of order 0.01 magnitudes of
inferred SFD extinction is sufficient to explain the anomaly.  Moreover,
simulation of the effect with mock data reproduces the anomaly's major
qualitative and quantitative features.

Assuming the above interpretation to be correct, our results represent
good news in one respect and bad news in another:

In the former sense, it is reassuring that the implied systematic errors
in the widely used SFD extinction map are quite small in magnitude, of
order hundredths of a magnitude, and in the range those authors
expected; in other words, they are not likely to be a significant
problem for most applications.

However, the bad news is that these systematic errors are themselves
correlated with the spatial clustering and distribution of galaxies in
some complex and potentially pernicious way.  For ``precision cosmology''
applications that depend sensitively on the accuracy of statistical
measures of galaxy clustering ({\it e.g.}, the power spectrum or baryon
acoustic oscillations), it will be necessary to disentangle the signal
from a systematic source of noise (the extinction) which depends on the
signal one is trying to measure \citep{Yahata2005,Eisenstein2005}.  
Moreover, studies of galaxy clustering
are ordinarily carried out in regions of the sky selected to have low
extinction, {\it i.e.,} just those in which the systematic extinction
errors are most strongly correlated with the large scale structure.  We
plan to address this and related problems in future work.

\bigskip
%%%%%%%%%%%%%%%%%%%%%%%%%%%%%%%%%%%%%%%%%%%%%%%%%%%%%%%%
We thank Naoki Yasuda and J. Mohr for useful discussions.  K. Y. was
supported by Grants-in-Aid for Japan Society for the Promotion of
Science Fellows.  ELT's participation in the project was supported in
part by NASA grant NAG5-13148.

Funding for the SDSS and SDSS-II has been provided by the Alfred P.
Sloan Foundation, the Participating Institutions, the National Science
Foundation, the U.S. Department of Energy, the National Aeronautics and
Space Administration, the Japanese Monbukagakusho, the Max Planck
Society, and the Higher Education Funding Council for England.  The SDSS
Web Site is http://www.sdss.org/.

The SDSS is managed by the Astrophysical Research Consortium for the
Participating Institutions. The Participating Institutions are the
American Museum of Natural History, Astrophysical Institute Potsdam,
University of Basel, Cambridge University, Case Western Reserve
University, University of Chicago, Drexel University, Fermilab, the
Institute for Advanced Study, the Japan Participation Group, Johns
Hopkins University, the Joint Institute for Nuclear Astrophysics, the
Kavli Institute for Particle Astrophysics and Cosmology, the Korean
Scientist Group, the Chinese Academy of Sciences (LAMOST), Los Alamos
National Laboratory, the Max-Planck-Institute for Astronomy (MPIA), the
Max-Planck-Institute for Astrophysics (MPA), New Mexico State
University, Ohio State University, University of Pittsburgh, University
of Portsmouth, Princeton University, the United States Naval
Observatory, and the University of Washington.

%%%%%%%%%%%%%%%%%%%%%%%%%%%%%%%%%%%%%%%%%% 
%
%  citation
%
%%%%%%%%%%%%%%%%%%%%%%%%%%%%%%%%%%%%%%%%%% 

%%%%%%%%%%%%%%%%%%%%%%%%%%%%%%%%%%%%%%%%%%%%%%%%%%%%%%%%%%%%%%%%%%%
% Figure 1
%%%%%%%%%%%%%%%%%%%%%%%%%%%%%%%%%%%%%%%%%%%%%%%%%%%%%%%%%%%%%%%%%%%
%%%%%%%%%%%%%%%%%%%%%%%%%%%%%%%%%%%%%%%%%%%%%%%%%%%%%%%%
% Figure 1 END
%%%%%%%%%%%%%%%%%%%%%%%%%%%%%%%%%%%%%%%%%%%%%%%%%%%%%%%%

%%%%%%%%%%%%%%%%%%%%%%%%%%%%%%%%%%%%%%%%%%%%%%%%%%%%%%%%
% Figure 2
%%%%%%%%%%%%%%%%%%%%%%%%%%%%%%%%%%%%%%%%%%%%%%%%%%%%%%%%
%%%%%%%%%%%%%%%%%%%%%%%%%%%%%%%%%%%%%%%%%%%%%%%%%%%%%%%%
% Figure 2 END
%%%%%%%%%%%%%%%%%%%%%%%%%%%%%%%%%%%%%%%%%%%%%%%%%%%%%%%%
%%%%%%%%%%%%%%%%%%%%%%%%%%%%%%%%%%%%%%%%%%%%%%%%%%%%%%%% 
% Figure 3
%%%%%%%%%%%%%%%%%%%%%%%%%%%%%%%%%%%%%%%%%%%%%%%%%%%%%%%%
%%%%%%%%%%%%%%%%%%%%%%%%%%%%%%%%%%%%%%%%%%%%%%%%%%%%%%%%
% Figure 3 END
%%%%%%%%%%%%%%%%%%%%%%%%%%%%%%%%%%%%%%%%%%%%%%%%%%%%%%%%
%%%%%%%%%%%%%%%%%%%%%%%%%%%%%%%%%%%%%%%%%%%%%%%%
% Figure 4
%%%%%%%%%%%%%%%%%%%%%%%%%%%%%%%%%%%%%%%%%%%%%%%%%
%%%%%%%%%%%%%%%%%%%%%%%%%%%%%%%%%%%%%%%%%%%%%%%%%%%%%%%
% Figure 4 END
%%%%%%%%%%%%%%%%%%%%%%%%%%%%%%%%%%%%%%%%%%%%%%%%%%%%%%%%
%%%%%%%%%%%%%%%%%%%%%%%%%%%%%%%%%%%%%%%%%%%%%%%%%%%%%%%% 
% Figure 5
%%%%%%%%%%%%%%%%%%%%%%%%%%%%%%%%%%%%%%%%%%%%%
%%%%%%%%%%%%%%%%%%%%%%%%%%%%%%%%%%%%%%%%%%%%%%%%%%%%%%%%
% Figure 7 END
%%%%%%%%%%%%%%%%%%%%%%%%%%%%%%%%%%%%%%%%%%%%%%%%%%%%%%%%
%%%%%%%%%%%%%%%%%%%%%%%%%%%%%%%%%%%%%%%%%%%%%%%%%%%%%%%%
% Figure 8
%%%%%%%%%%%%%%%%%%%%%%%%%%%%%%%%%%%%%%%%%%%%%%%%%%%
%%%%%%%%%%%%%%%%%%%%%%%%%%%%%%%%%%%%%%%%%%%%%%%%%%%%%%%%
% Figure 8 END
%%%%%%%%%%%%%%%%%%%%%%%%%%%%%%%%%%%%%%%%%%%%%%%%%%%%%%%%
%%%%%%%%%%%%%%%%%%%%%%%%%%%%%%%%%%%%%%%%%%%%%%%%%%%%%%%% 
% Figure 9
%%%%%%%%%%%%%%%%%%%%%%%%%%%%%%%%%%%%%%%%%%%%%%%%
%%%%%%%%%%%%%%%%%%%%%%%%%%%%%%%%%%%%%%%%%%%%%%%%%%%%%%%%
% Figure 9 END
%%%%%%%%%%%%%%%%%%%%%%%%%%%%%%%%%%%%%%%%%%%%%%%%%%%%%%%%

%%%%%%%%%%%%%%%%%%%%%%%%%%%%%%%%%%%%%%%%%%%%%%%%%%%%%%%%
% Figure 10
%%%%%%%%%%%%%%%%%%%%%%%%%%%%%%%%%%%%%%%%%%%%%%%%%%%
%%%%%%%%%%%%%%%%%%%%%%%%%%%%%%%%%%%%%%%%%%%%%%%%%%%%%%%%
% Figure 10 END
%%%%%%%%%%%%%%%%%%%%%%%%%%%%%%%%%%%%%%%%%%%%%%%%%%%%%%%%
%%%%%%%%%%%%%%%%%%%%%%%%%%%%%%%%%%%%%%%%%%%%%%%%%%%%%%%%
% Figure 11
%%%%%%%%%%%%%%%%%%%%%%%%%%%%%%%%%%%%%%%%%%%%%%%
%%%%%%%%%%%%%%%%%%%%%%%%%%%%%%%%%%%%%%%%%%%%%%%%%%%%%%%%
% Figure 11 END
%%%%%%%%%%%%%%%%%%%%%%%%%%%%%%%%%%%%%%%%%%%%%%%%%%%%%%%%

%%%%%%%%%%%%%%%%%%%%%%%%%%%%%%%%%%%%%%%%%%%%%%%%%%
% Figure 12
%%%%%%%%%%%%%%%%%%%%%%%%%%%%%%%%%%%%%%%%%%%%%%%%%%
%%%%%%%%%%%%%%%%%%%%%%%%%%%%%%%%%%%%%%%%%%
% Figure 12 END
%%%%%%%%%%%%%%%%%%%%%%%%%%%%%%%%%%%%%%%%%%
%%%%%%%%%%%%%%%%%%%%%%%%%%%%%%%%%%%%%%%%%%
% Figure 13
%%%%%%%%%%%%%%%%%%%%%%%%%%%%%%%%%%%%%%
%%%%%%%%%%%%%%%%%%%%%%%%%%%%%%%%%%%%%%%%%%
% Figure 13 END
%%%%%%%%%%%%%%%%%%%%%%%%%%%%%%%%%%%%%%%%%%
%%%%%%%%%%%%%%%%%%%%%%%%%%%%%%%%%%%%%%%%%% 
% Figure 14
%%%%%%%%%%%%%%%%%%%%%%%%%%%%%%%%%%%%%%
%%%%%%%%%%%%%%%%%%%%%%%%%%%%%%%%%%%%%%%%%%
% Figure 14 END
%%%%%%%%%%%%%%%%%%%%%%%%%%%%%%%%%%%%%%%%%%
\end{document}